\documentclass[aip,rsi,amsmath,amssymb,reprint]{revtex4-1}

\usepackage{graphicx,caption,subcaption}
\usepackage{hyperref}
\usepackage[capitalise]{cleveref}
\usepackage[normalem]{ulem} 
\usepackage{xcolor} 

\begin{document}

\title{Dimerization of dehydrogenated polycyclic aromatic hydrocarbons on graphene} 

\author{Zeyuan Tang}
\affiliation{Center for Interstellar Catalysis, Department of Physics and Astronomy, Aarhus University, Ny Munkegade 120, Aarhus C 8000, Denmark}

\author{Bjørk Hammer}
\email{hammer@phys.au.dk}
\affiliation{Center for Interstellar Catalysis, Department of Physics and Astronomy, Aarhus University, Ny Munkegade 120, Aarhus C 8000, Denmark}


\begin{abstract}
Dimerization of polycyclic aromatic hydrocarbons (PAHs) is an important, yet poorly understood step in on-surface synthesis of graphene (nanoribbon), soot formation, growth of carbonaceous dust grain in the interstellar medium (ISM).
The on-surface synthesis of graphene and the growth of carbonaceous dust grain in the ISM require the chemical dimerization in which chemical bonds are formed between PAH monomers.
An accurate and cheap method of exploring structure rearrangements is needed to reveal the mechanism of chemical dimerization on surfaces.
This work has investigated the chemical dimerization of two dehydrogenated PAHs (coronene and pentacene) on graphene via an evolutionary algorithm augmented by machine learning surrogate potentials and a set of customized structure operators.
Different dimer structures on surfaces have been successfully located by our structure search methods.
Their binding energies are within the experimental errors of temperature programmed desorption (TPD) measurements.
The mechanism of coronene dimer formation on graphene is further studied and discussed.
\end{abstract}

\maketitle

\section{Introduction}
Dimerization of polycyclic aromatic hydrocarbons (PAHs) plays an important role in on-surface synthesis of graphene (nanoribbon)\cite{wang2017j,chen2020c}, soot formation\cite{kholghy2018,mercier2019,frenklach2020}, and growth of dust grain in the interstellar medium .\cite{contreras2013,fulvio2017,thomas2020,marin2020}
There are two types of PAH dimerization depending on how PAHs are linked together.
One is physical dimerization where PAHs are combined through weak van der Waals interactions.
The other one is chemical dimerization where covalent bonds are formed between PAHs.
However, a comprehensive understanding of dimerization of PAHs has not been reached yet.
Previous theoretical attempts to mimic gas-phase combustion chemistry include molecular dynamics simulations on binary PAH collisions\cite{schuetz2002,mao2017}, quantum chemical studies on interactions between PAHs\cite{zhang2014f}.
PAH dimerization on surfaces (e.g. self-assembly of PAH on graphite \cite{florio2005}) is an important ingredient in soot particle and dust grain chemistry as well.
Recently, the chemical dimerization of dehydrogenated PAHs on graphite has been studied using laser desorption coupled with ion mobility spectrometry\cite{weippert2020}.
In this work, some possible dimer structures are proposed by gas phase density functional theory (DFT) calculations.
To ascertain the dimer structure, one needs to compare the binding energy of dimers on surface from theory and temperature programmed desorption.
Meanwhile, the dimer formation mechanism on surfaces remains unknown.
Many factors are needed to be considered in order to understand PAH dimerization and growth.
For example, temperature and PAH mass can determine whether physical or chemical growth is favorable and have a significant impact on the size of the formed soot particles. \cite{mao2017a}
The growth of small PAHs are also dependent on temperature and pressure. \cite{mebel2017a}
When it comes to the dynamics and kinetics of PAH dimerization, PAH collision settings such as impact parameters and orientations have a huge influence on the lifetime of the formed PAH dimers and the reaction rates. \cite{mao2017}
For the PAH growth in the interstellar medium, ultraviolet and cosmic-ray irradiations cannot be ignored. \cite{fulvio2017}

Finding energetically favorable structures is often the first step in studying chemical reactions before more kinetic or dynamical analysis is performed.
In the case of the chemical PAH dimerization, resolving the complex chemical bond rearrangements is necessary and relies heavily on efficient structure search methods.
We only focus on the structure search and ignore other factors affecting the chemical PAH dimerization in this study.
A global exploration scheme combining parallel tempering Monte Carlo and local quenches was used to explore stacking patterns and search stable structures of pyrene clusters.\cite{dontot2019}
However, few efforts have been dedicated to the chemical PAH dimerization using global optimization techniques because the structure complexity is increased from physical dimerization to chemical dimerization.
The evolutionary algorithm (EA) approach\cite{oganov2006}, another
popular global optimization scheme, has been successfully applied to a
wide range of systems including bulk crystals,\cite{oganov2006}
surface reconstructions,\cite{sorensen2018} and supported nanoclusters \cite{vilhelmsen2014}.
It has also been extended to molecular crystals\cite{zhu2012} and
polymers\cite{zhu2014a} where the mutations in the form of structure operators are imposed on a group of atoms instead of individual atoms.
Such an extension assumes that the bond connectivity within monomers is fixed.
This assumption may not be valid for the chemical dimerization of PAHs due to possible hydrogen migration.
Therefore, it is necessary to introduce a new set of structure operators which allow the bond connectivity within monomers to change.

In this paper, we propose a new set of structure operators within the
EA framework, which fixes the carbon skeleton but allows hydrogen
atoms to jump between carbon atoms.
PAH dimers on graphene often contain hundreds of atoms in a large supercell \cite{graphite_coronene} and require vdW functionals\cite{vdW-DF,vdW-DF2,klimes2009} to accurately describe the interaction between PAHs and graphene.\cite{thrower2013,berland2013}
The requirements of both large system size and vdW functional render the calculations extremely expensive especially in an EA search.
The global optimization with first-principles energy expressions
(GOFEE) \cite{GOFEE} is a promising method to reduce computational
costs by replacing the local relaxation in the first principles potential with the local relaxation in a gaussian process regression model.
GOFEE further produces a number of such new candidate structures and
selects based on a Bayesian acquisition function only the most
promissing one for the expensive DFT evaluation.
Our work will combine the newly introduced structure operators and GOFEE to find stable PAH dimers on graphene.
Two dehydrogenated PAH systems (coronene and pentacene) are chosen by the inspiration from the experiment by Weippert et al. \cite{weippert2020}.

\section{Method}
This section describes the theoretical methods utilized in this work.
The structure search method is introduced first.
It is based on GOFEE and a set of newly introduced operators in \cref{s-operator}.
Afterwards, the setup for the structure search is described in \cref{s-structure}.
Finally, other computational details are mentioned in \cref{s-computation}.

\subsection{Operators} \label{s-operator}
\begin{figure*}
	\begin{center}
	\includegraphics[width=\textwidth]{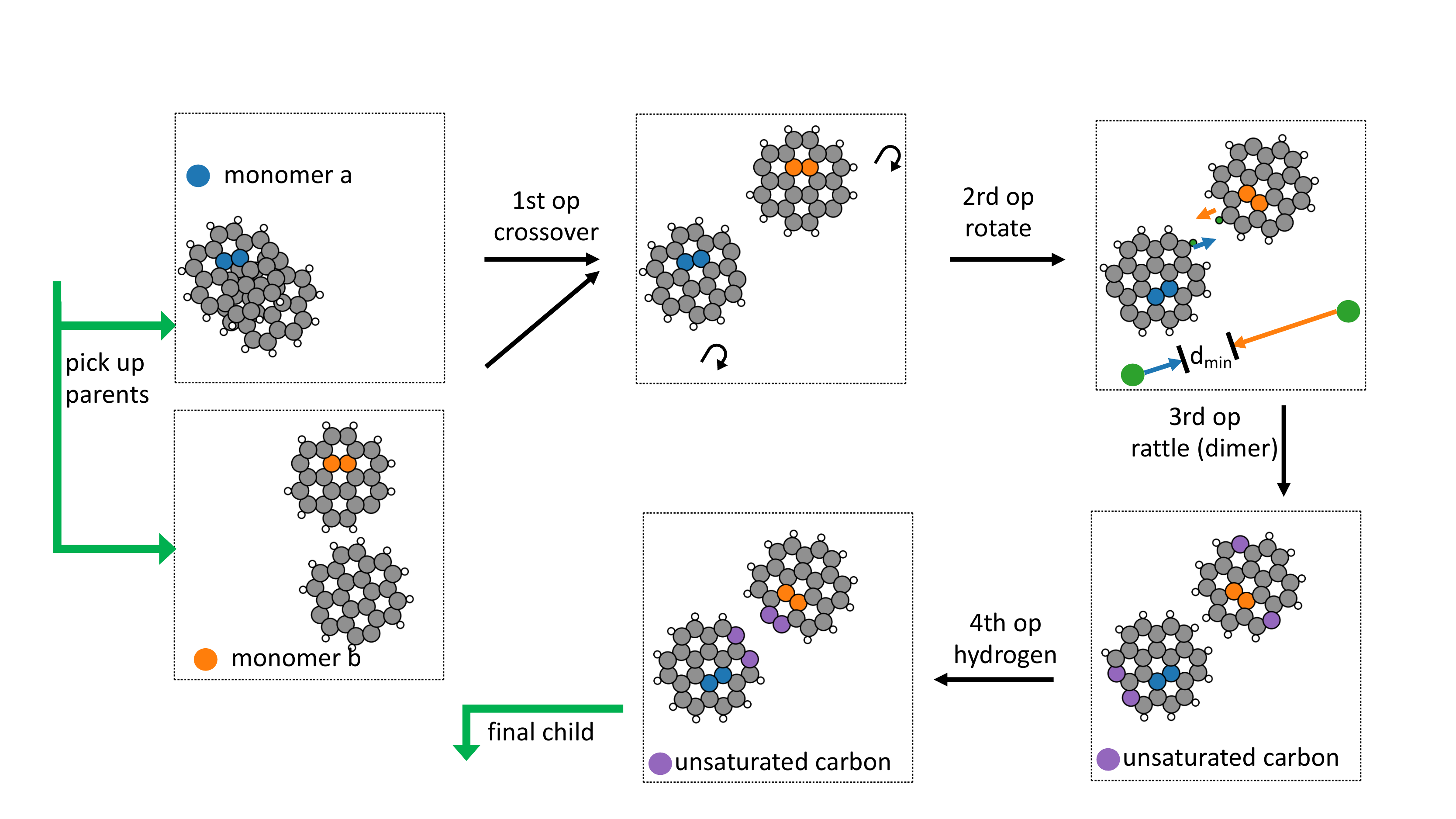}
	\caption{
        Scheme of structure operators for generating promising candidates of PAH dimer on graphene.
        At the beginning, two structures are randomly picked from the population.
        Half of each structure are paired later.
        Each monomer in the newly generated structure has rotational degrees of freedom to adjust their orientations.
        The third step tries to diminish the distance between two monomers. The closest atom pairs are marked in green.
        Finally, a candidate structure is sent for local relaxation after hydrogen vacancies are generated in the overlapping region of two monomers.
    }
    \label{fig:cGOFEE}
    \end{center}
\end{figure*}

There are several requirements for structure operators when searching for PAH dimers on graphene.
\begin{itemize}
    \item The structure generated by these operators should not be biased by the prior knowledge of dimer structures and adsorption sites.
    \item Operations need to be imposed on groups of atoms (treated as a monomer) with the aim of confining the search space and avoiding unnecessary exploration.
    In practice, a unique tag is assigned to each monomer to keep its identification during structure operations.
    This has previously been introduced in methods for predicting molecular crystals.\cite{zhu2012}
    \item The bond connectivity in monomers is partially fixed, meaning the carbon skeleton remains unchanged but the hydrogen atoms are allowed to move.
    Different types of dimers (being fused via 4, 5, 6-member rings) can be formed by having this partially flexible bond connectivity.
    We will design a hydrogen operator to address this issue.
    \item Structure operators should explore all possible adsorption configurations (top, bridge, hollow, \dots).
    \item Since the action of a single operation or several operations
      in combination may easily result in dimer structures where the monomers
      are too far away to interact or too close to form proper chemical bonds, 
      the structure operators must be imposed in a properly controlled
      sequence. This is illustrated in \cref{fig:cGOFEE}.
    The operators start by paring two random structures from the
    population according to the method detailed in \cref{s-crossover}.
    Then, the resulting structure would explore different orientations
    as described in \cref{s-rotation}.
    In order to avoid two monomers being too far, they are displaced
    according to the method explained in \cref{s-rattle}.
    Finally, hydrogen atoms in the overlapping region of two monomers
    are removed and reintroduced elsewhere as described in \cref{s-hydrogen}.
\end{itemize}
Based on these requirements, we have designed the following operators.

\subsubsection{Crossover} \label{s-crossover}
Unique tags are assigned to each monomer (one dehydrogenated PAH) for easy identification.
Only the center of mass and orientation of each monomer are used when paring two structures from the population with crossover operators.
\subsubsection{Rotation} \label{s-rotation}
Since the focus of this work is on-surface dimerization, the dominant rotation for each monomer is around the $z$-axis with angles randomly sampled from 0 to 180 degrees.
The rotations around the $x$ and $y$ axes are limited to a maximum angle of 15 degrees.
The center of mass of the monomers is chosen as the rotation center.
\subsubsection{Rattle} \label{s-rattle}
Unlike traditional rattle operators in EA, our rattle operator is not purely random, but tries to bring two separate monomers together.
The direction of the rattle motion is pointing from the center of mass
of one monomer towards that of the other.
The distance through which each monomer can move is random, but the distance between the closest atoms ($d_{min}$ in \cref{fig:cGOFEE}) after the move is set to 1.5Å.
\subsubsection{Hydrogen} \label{s-hydrogen}
This is a key operator for generating promising dimer structures.
The first step is to locate the (first, second, third, etc.) closest pairs of edge carbon atoms (one carbon in one monomer, another carbon in another monomer).
Then hydrogen atoms are removed from these carbon sites.
The final step is to bind deleted hydrogen atoms on some edge carbon atoms.
If an edge carbon atom has not been linked to any hydrogen atoms yet, hydrogen atoms will preferentially be placed in a proper distance first and in the same plane as the monomer.
Once all edge carbon atoms are occupied by hydrogen atoms, any remaining hydrogen atoms will be placed randomly on top of edge carbon atoms.

\subsection{Structure search} \label{s-structure}
Initial structures are generated by placing two separate monomers randomly on the surface of graphene.
The number of hydrogen vacancies were initially set to 2 and 4 for the coronene monomer and the pentacene monomer to be consistent with the chemical formulas of the coronene dimer ([Cor$-$2H]$_2$) and the pentacene dimer ([Pen$-$4H]$_2$) in Weippert et al. \cite{weippert2020}.
During the search of the pentacene dimer, we found the configuration in with 5 C-C bonds connect two monomers had much lower energies than the one with 4 connecting C-C bonds.
Therefore, the number of hydrogen vacancies for the pentancene monomer was switched to 5.
The number of independent runs was set to 20 and 10 for gas phase and on-surface search respectively.
One run is counted as a success in the gas phase search if the total energy is within 0.1 eV of the most stable structure.
The energy threshold for defining the success in the on-surface search can be a little higher, but the adsorption configuration must match with the one in the most stable structure by comparing their stacking modes.

\subsection{Computational setup} \label{s-computation}
The Atomic Simulation Environment (ASE) \cite{ASE} was used to set up atomistic structures.
Structure searches in the gas phase were conducted within the framework of linear combination of atomic orbitals (LCAO) \cite{larsen2009} in in GPAW v19.8.1 \cite{GPAW}.
The PBE functional \cite{PBE} and double-zeta polarized basis set were used in all LCAO calculations.
A $10\times 10$ graphene monolayer was used for on-surface structure searches.
All on-surface calculations were performed using the real-space, grid-based projector augmented wave method in GPAW.
The grid spacing was 0.20 Å.
The opt-PBE vdW functional \cite{klimes2009} was used because it
offers an accurate description of the interaction between PAHs and graphene \cite{thrower2013}.
It also has a good agreement with G3(MP2,CC) (a high-level ab initio method) for the H-migration barrier of selected PAH systems \cite{kislov2005,mebel2017a}, see Fig. S3 and Fig. S4 of the supplementary material.
The structures found by GOFEE were relaxed in DFT until the maximum force reached 0.02 eV/Å.
The adsorption energies were calculated using the relaxed structures and compared with experiments.
The transition states were located through the AutoNEB \cite{AutoNEB} which inserts images between initial and final states on the fly and adopts a modified spring force.
Three images were running simultaneously and no more images were inserted when the accumulated images reached a maximum number (10 $\sim$ 20 images depending on system).
The force convergence criterion on transition states was the same as for local minima.
All structure relaxations and NEB calculations were done with spin paired settings.
Spin polarization was only applied in single point calculations of optimized structures.
Therefore, all reported adsorption energies and energy barriers would contain contributions from the spin state.
More discussions on the spin state can be found in the supplementary material.

\section{Results}
In this study, two dehydrogenated polycyclic aromatic hydrocarbons (coronene and pentacene, with two and four hydrogen atoms missing respectively) \cite{weippert2020} were used as monomers for dimerization on graphene.
\subsection{Coronene}
\begin{figure}
	\centering
	\includegraphics[width=1\linewidth]{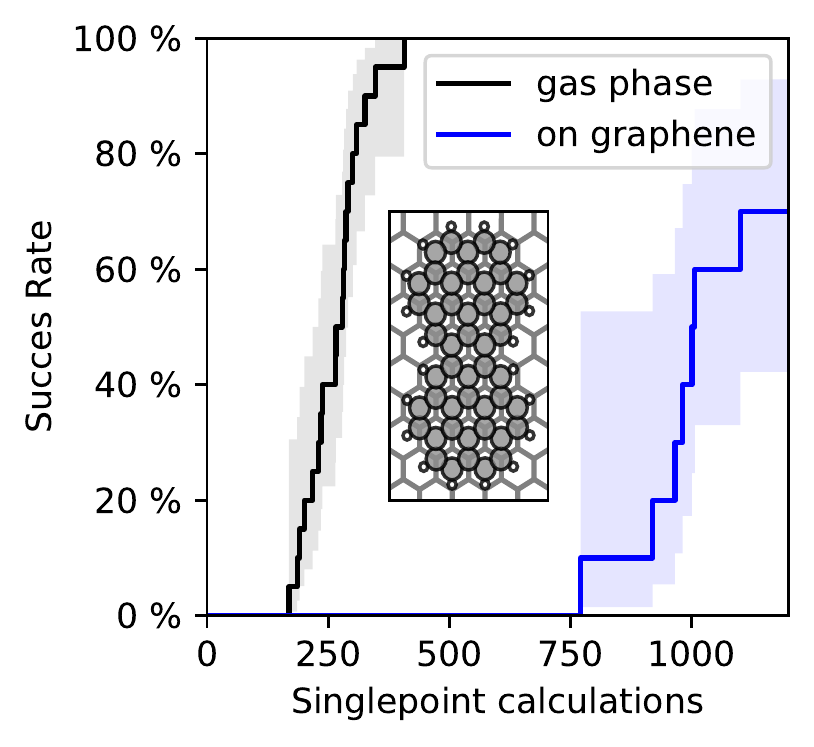}
	\caption{
        Success rate of coronene dimer search in the gas phase (black line) and on graphene (blue line).
        The inserted image shows the top view of the global minimum of the coronene dimer on graphene.
    }
	\label{fig:cor-success}
\end{figure}

Coronene is the first system for testing our structure search method.
The search begins with two coronene monomers in which two hydrogen atoms are missing.
\cref{fig:cor-success} shows the success rate of the coronene dimer search in the gas phase and on graphene.
It is easy to find the most stable configuration of the coronene dimer ([Cor$-$2H]$_2$) in the gas phase.
All 20 runs have successfully found the global minimum using less than 400 single point calculations.
Stacked or other planar configurations did show up during the search.
However, they would be quickly overtaken by the planar dimer configuration with 6 membered ring in the center (shown in the insert of \cref{fig:cor-success}), as evidenced by rapid increases in the success rate.
The search on graphene is more difficult, since the method needs to not only find the most stable dimer structure, but also locate the most favorable adsorption site.
Some of the high-energy structures found during the search on graphene are given in Fig. S1.
The larger search space on graphene also means that more single point
calculations are needed than for the gas phase search.
The extra computational time was mainly spent on exploring adsorption sites, while the planar dimer configuration with 6 membered ring in the center has been located in the early stage of searches.
In order to improve the success rate of the dimer search on graphene, there is definitely room for optimizing hyperparameters such as rotation angles, rattle amplitudes and ratios between different operators.
However, those hyperparameters are very likely to be system dependent.
Our study is a proof-of-concept work to show GOFEE with modified structure operators is capable of locating PAH dimers on surfaces.
Therefore, the fine tuning of hyperparameters was not performed.

\begin{figure}
	\centering
	\includegraphics[width=0.8\linewidth]{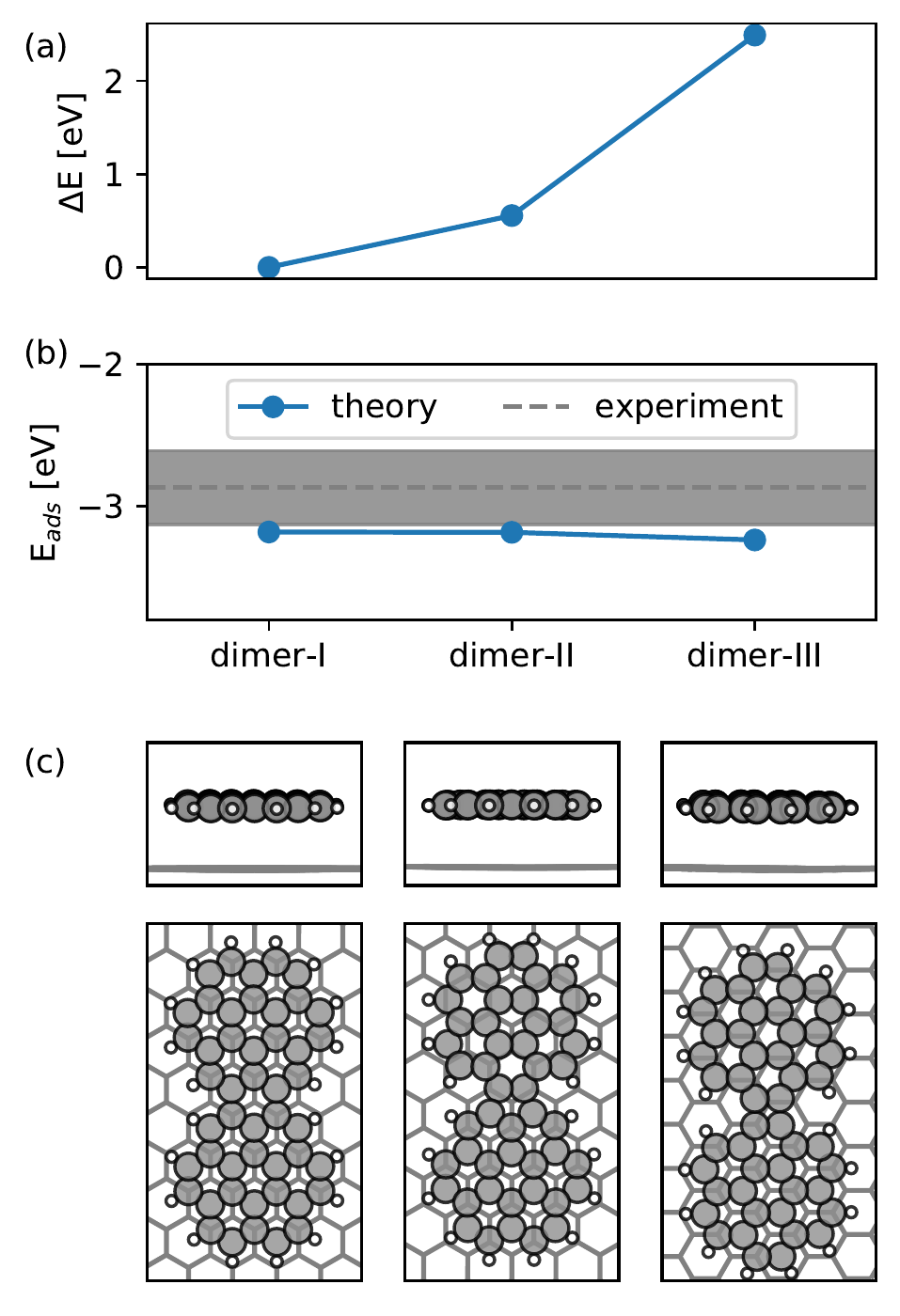}
	\caption{
        Adsorption of different coronene dimers on graphene: 
        (a) Relative energies with respect to the most stable dimer. 
        (b) Computed adsorption energies vs experimental values from Weippert et al. \cite{weippert2020}.
        For the experimental values, the mean is plotted in a gray dashed line while the derivation is plotted in gray shaded region.
        (c) Side and top views of adsorption configurations.
    }
    \label{fig:cor-Eads}
\end{figure}
The comparison between the theoretical adsorption energy for the global minimum is given in \cref{fig:cor-Eads}.
Another two low-energy configurations (dimer-II and dimer-III) are included as well.
Theoretical adsorption energies for these three configurations are all within the range of experiments, thus not distinguishable.
Due to the lowest relative energy, the dimer with a 6-membered ring (dimer-I, also called dicoronylene (DCY)) is the most plausible dimer structure.
This agrees well with experiments where DCY and products from ion deposition yield identical binding energies on highly oriented pyrolytic graphite (HOPG).\cite{weippert2020}

\begin{figure*}
	\centering
	\includegraphics[width=1\textwidth]{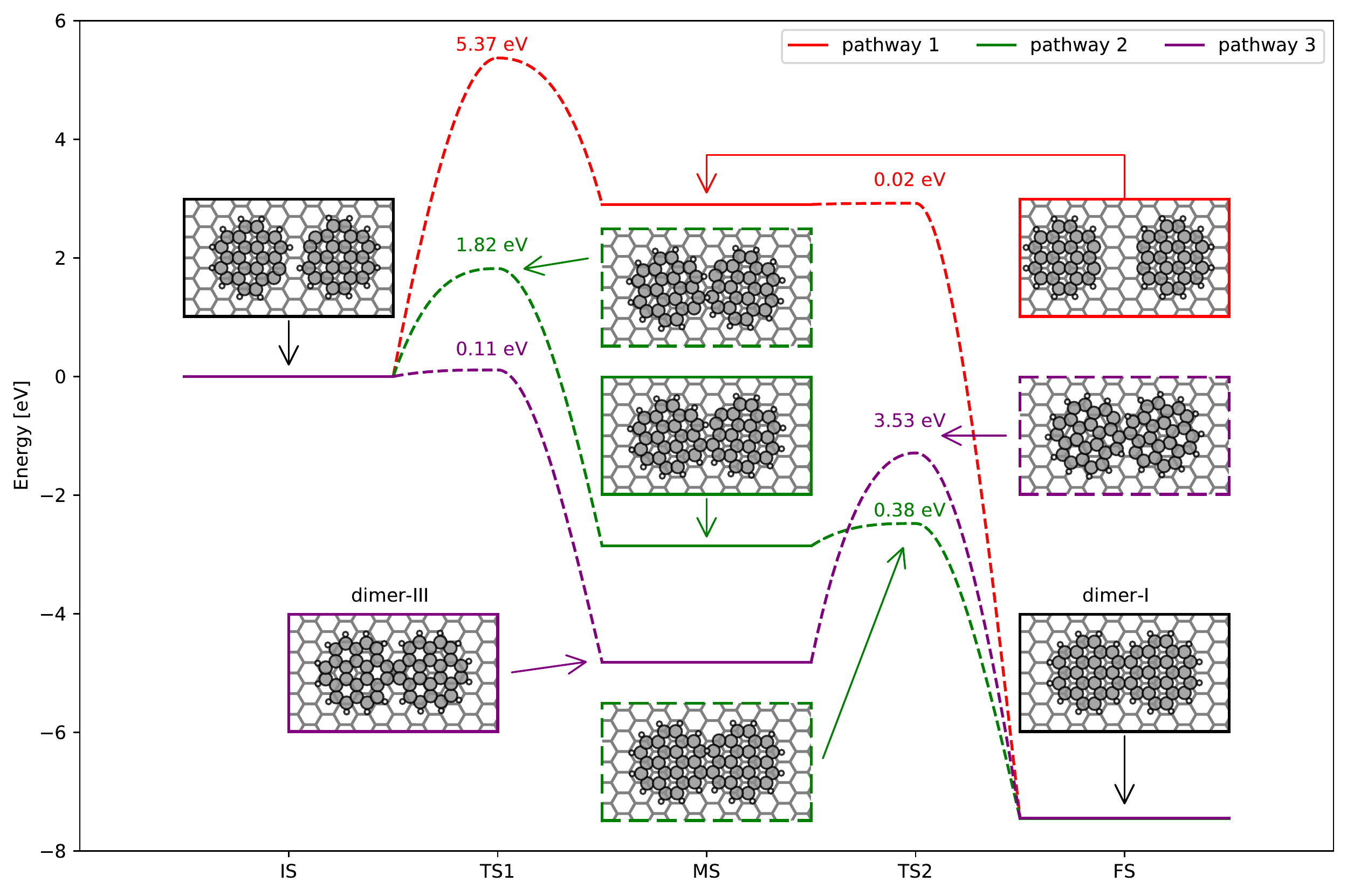}
	\caption{
        Energy profile of dimer formation from two separate monomers on graphene.
        Different pathways are shown in different colors.
        The solid lines represent local minima (IS: initial state; MS: meta state; FS: final state).
        Transition states (TS) are plotted in dashed lines with energy barriers attached.
    }
	\label{fig:cor-neb}
\end{figure*}
The formation of dimer-I is important for understanding the dimerization of dehydrogenated coronene on graphene.
Two doubly dehydrogenated coronene monomers (referred to as:
[Cor$-$2H]) are chosen as the start of the dimerization, because
it is possible to obtain the [Cor$-$2H] monomer from the dehydrogenation of pristine coronene by electron impact experiments \cite{weippert2020}.
The [Cor$-$2H] monomer can be mass-selected and deposited on graphene.
The reaction from two [Cor$-$2H] monomers to [Cor$-$2H]$_2$ is known to occur based on the experimentally available mass information.
There may exist other mechanisms for the formation of [Cor$-$2H]$_2$ if the start of the reaction is pristine coronene.
Those possibilities are not explored here, since the focus of this work is the dimerization from two [Cor$-$2H] monomers to the most stable configuration of [Cor$-$2H]$_2$, dimer-I.
\cref{fig:cor-neb} shows three routes to convert two monomers into the dimer-I.
(1) The monomer with armchair dehydrogenated sites is transformed to the one with zigzag dehydrogenated sites via the hydrogen migration which needs to overcome a 5.37 eV barrier.
The hydrogen migration contains some separate steps and is described in details in Fig. S6.
The translation and rotation of both types of monomers require about
0.1 eV kinetic energy.
Such a low barrier increases the chance for two monomers to meet with each other.
The combination of two monomers with zigzag dehydrogenated sites is even easier than the translation and rotation of the monomer.
(2) The hydrogen transfer within the monomer is happening simultaneously with the dimer formation.
The first step in this path is that one of the hydrogen atoms in between two monomers migrates from one monomer to the other.
The hydrogen atom in the transition state is positioned on the bridge site of the newly formed C-C bond which connects two monomers.
Then the hydrogen atom jumps to the unsaturated carbon in the left monomer.
The formation of the second C-C bond involves a similar hydrogen
transfer mechanism, but only needs to overcome a 0.38 eV barrier which
is significantly lower than the 1.82 eV in the first C-C bond formation.
Detailed energy profiles and intermediate structures of the first and second C-C bond formation are given in Fig. S8 and Fig. S9.
(3) Two monomers are combined directly under the overall translation and rotation on graphene.
The formed dimer with a 4-member ring is less stable than the one with 6-member ring and requires 3.53 eV energy to achieve this conversion.
Such a high barrier implies that it is hard to open a ring once it is formed, even though the coronene dimer with a 4-member ring has much a higher energy than that with the 6-member ring.

The presented computational results allow for an assessment of experimentally derived mechanisms.
It has been suggested that the dimerization is promoted by ion diffusion on surface.\cite{weippert2020}
The calculated barriers for translation and rotation of monomers on graphene are around 0.1 eV, making such fast diffusion feasible.
Another requirement of [Cor$-$2H]$_2$ formation is the hydrogen transfer between monomers.\cite{weippert2020}
Inspecting the energy profiles in \cref{fig:cor-neb}, it is seen that the pathway leading to the thermodynamically preferred dimer, dimer-I, is hindered by a large barrier of 1.82 eV.
Instead, the energy profiles reveal facile formation of the less stable dimer, dimer-III.
Further isomerization of this isomer into dimer-I is not likely as the calculated energy profiles show a 3.53 eV barrier for that.
While it cannot be ruled out that other lower energy barrier pathways exist for such a transformation, our results indicate that the formation of dimer-I on graphene will be kinetically hindered and that dimer-III will not.
This is not at variance with the available experimental information that is merely based on the mass of the dimer formed.
We cannot exclude the possibilities of collision-driven dimerization \cite{mao2017} because kinetic energies of monomers are not considered in our calculations.
It remains unclear whether the dimerization occurs during/after the deposition or during the temperature programmed desorption.
The contributions from positively charged species are not covered in this study.

\subsection{Pentacene}
\begin{figure}
	\centering
	\includegraphics[width=1\linewidth]{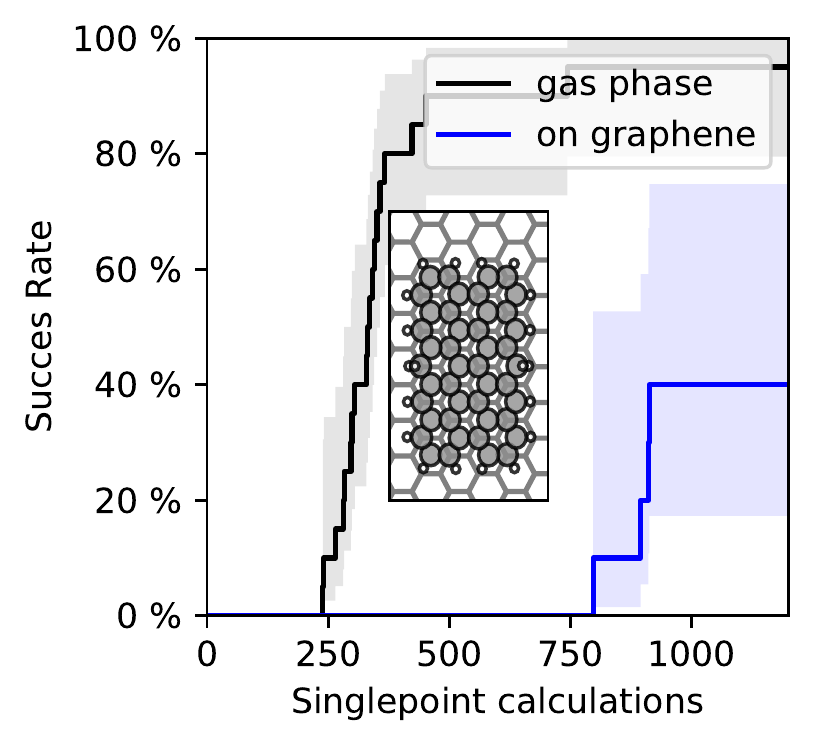}
	\caption{
        Success rate of pentacene dimer search in the gas phase (black line) and on graphene (blue line).
        The inserted image shows the top view of the global minimum of the pentacene dimer on graphene.
    }
	\label{fig:pen-success}
\end{figure}

\begin{figure*}
	\begin{center}
	\includegraphics[width=0.8\textwidth]{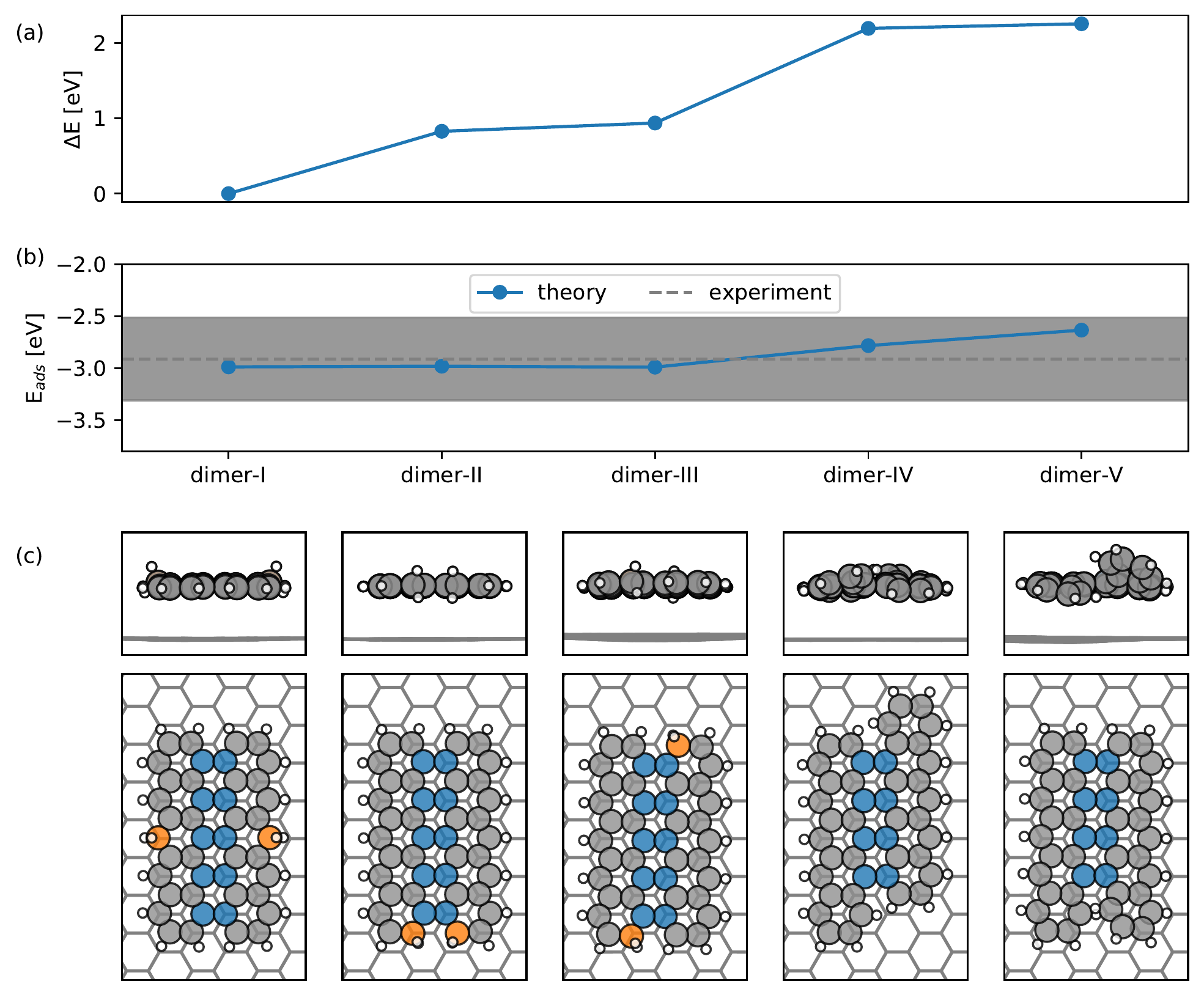}
	\caption{
        Adsorption of different pentacene dimers on graphene: 
        (a) Relative energies with respect to the most stable dimer.
        (b) Adsorption energies vs experimental values from Weippert et al. \cite{weippert2020}.
        For the experimental values, the mean is plotted in a gray dashed line while the derivation is plotted in gray shaded region.
        (c) Side and top views of adsorption configurations.
        The carbon atoms connecting two monomers are highlighted in blue while the carbon atoms with excess hydrogen atoms are shown in orange.
    }
    \label{fig:pen-Eads}
    \end{center}
\end{figure*}

Pentacene is less symmetric than coronene and the monomer used in the search has four hydrogen vacancies.
Therefore, the pentacene dimer experiences a more diverse configuration space.
The structure diversity is reflected in the need for more single point
calculations to achieve the same success rate as for the coronene dimer in the gas phase.
It makes the dimer search on graphene even harder and the success rate
saturates at around 40\%, cf.\ \cref{fig:pen-success}.
The global minimum shown in \cref{fig:pen-success} (same as dimer-I in \cref{fig:pen-Eads}) has five C-C bonds connecting two monomers while each monomer only has four hydrogen vacancies.
Weippert et al. \cite{weippert2020} has proposed the dimers (dimer-IV and dimer-V in \cref{fig:pen-Eads}) in which the connecting unit has four C-C bonds.

To fully characterize the possible structure of the pentacene dimer on graphene,
the global minimum energy structure and some four other highly stable
structures found during the EA search were subsequently relaxed in DFT
without constraints. Thereby the optimized structures were
obtained. These are depicted in \cref{fig:pen-Eads} (c) that shows how
the pentacene dimer in all five cases adopts the same type of hollow
adsorption site configuration.
The adsorption energies of the five dimers shown in
\cref{fig:pen-Eads} (b) all lie in the range of experimental values
and the five dimers are therefore equally good structural candidates
for the observed pentacene dimers.
Considering the relative energies, dimer-I, dimer-II, dimer-III with five connecting C-C bonds (blue parts in \cref{fig:pen-Eads} (c)) are more stable than dimer-IV, dimer-V with four connecting C-C bonds.
Forming one more C-C bond appears to lower the total energy at least 1 eV.
The finding of dimer-I, dimer-II, dimer-III is unexpected since dimer-IV and dimer-V are more close to chemical intuition and proposed by the experimental work \cite{weippert2020}.
In other words, it proves that our EA based search method is capable of predicting low energy dimers in an unbiased and precise fashion.
Meanwhile, the structure of the global minimum, dimer-I indicates the special hydrogen affinity of central edge carbon atoms in pentacene \cite{campisi2020}.

The pentancene dimer formation is more complex than the coronene case.
If the dimerization starts from two monomers with four hydrogen atoms missing on the long edge of pentacene, the direct combination of the two monomers will lead to either dimer-IV or dimer-V.
The conversion from dimer-IV to the global minimum dimer-I seems impossible because it involves the breaking of several C-C bonds.
The hydrogen migration might be feasible during the transition from dimer-V to dimer-I, but the hydrogen atoms would need to travel via many carbon sites.
Another problem is that the most stable monomer is the one with four hydrogen vacancies on the same outer ring according to our test.
It has also been observed in experiments that dehydrogenated pentacene is able to rehydrogenate again on Ir(111) \cite{curcio2021}.
Considering the need for considerable hydrogen migration, we do,
however, expect that the formation of the global minimum energy
structure will be kinetically hindered.

\section{Conclusion}
We have demonstrated that GOFEE with modified structure operators is capable of locating energetically favorable PAH (i.e. coronene and pentacene) dimer structures on graphene.
This method can, in principle, be applied to the dimerization of arbitrary PAH on surfaces.
The adsorption energies of the obtained PAH dimers on graphene are computed and compared with experiments.
Different isomers of PAH dimer have quite close adsorption energies with 0-0.5 eV derivations even though their relative energies differ by $\sim$2 eV.
It remains challenging to identify the dimer structure by comparing the adsorption energies of different isomers with experiment.
We have further studied the mechanism of coronene dimer formation on graphene.
It has been found that the formation of the most thermodynamically stable dimer is kinetically hindered.
More investigations are needed to provide valuable information on how the chemical dimerization of PAHs proceeds on graphene.

\section*{Supplementary Material}
See supplementary material for the high-energy structures found during the coronene dimer search and the discussion of the spin state.

\begin{acknowledgments}
This work has been supported by the Danish National Research Foundation through the Center of Excellence “InterCat” (Grant agreement no.: DNRF150), the European Union (EU) and Horizon 2020 funding award under the Marie Skłodowska-Curie action to the EUROPAH consortium, grant number 722346.
We acknowledge support from VILLUM FONDEN (Investigator grant, Project No. 16562).
\end{acknowledgments}

\bibliography{ref}

\end{document}